\documentclass[12pt,preprint]{aastex}
\usepackage{natbib}

\begin{document}
\title{Revisiting the Thermal Stability of Radiation-dominated Thin Disks}
\shorttitle{Thermal Stability of Thin Disks} \shortauthors{Zheng et
al.}
\author{Sheng-Ming Zheng\altaffilmark{1}, Feng Yuan\altaffilmark{2},
Wei-Min Gu\altaffilmark{1}, and Ju-Fu Lu\altaffilmark{1}}
\altaffiltext{1}{Department of Physics and Institute of Theoretical Physics and Astrophysics, Xiamen University, Xiamen, Fujian 361005, China}
\altaffiltext{2}{Key Laboratory for Research in Galaxies and Cosmology,
Shanghai Astronomical Observatory, Chinese Academy of Sciences, 80
Nandan Road, Shanghai 200030, China}
\email{guwm@xmu.edu.cn}
\begin{abstract}
The standard thin disk model predicts that when the accretion rate
is over a small fraction of the Eddington rate, which corresponds to
$L\ga 0.06 L_{\rm Edd}$, the inner region of the disk is
radiation-pressure-dominated and thermally unstable. However,
observations of the high/soft state of black hole X-ray binaries
with luminosity well within this regime ($0.01L_{\rm Edd}\la L\la
0.5L_{\rm Edd}$) indicate that the disk has very little variability,
i.e., quite stable. Recent radiation magnetohydrodynamic simulations
of a vertically stratified shearing box have confirmed the absence
of the thermal instability.
In this paper, we revisit the thermal stability by linear
analysis, taking into account the role of magnetic field in the
accretion flow.
By assuming that the field responses negatively to a positive
temperature perturbation, we find that the threshold of accretion
rate above which the disk becomes thermally unstable increases
significantly compared with the case of not considering the role of
magnetic field. This accounts for the stability of the observed
sources with high luminosities. Our model also presents a possible
explanation as to why only GRS 1915+105 seems to show thermally
unstable behavior. This peculiar source holds the highest accretion
rate (or luminosity) among the known high state sources, which is
well above the accretion rate threshold of the instability.
\end{abstract}
\maketitle

\section{Introduction}

The standard thin disk is a milestone in the development of
accretion disk theory. It successfully explains many observations
such as the ``big blue bump'' in active galactic nuclei (see review
in Frank et al. 2002) and the spectrum of high/soft state of black
hole X-ray binaries (Gierli\'nski \& Done 2004; hereafter GD04).
However, some questions still remain unsolved (e.g., Koratkar \&
Blaes 1999). One of them, for example, is how to explain the
observed hard X-ray emission in black hole sources. The temperature
of the thin disk is too low to produce the X-ray emission.

In this paper we will discuss another puzzle, which is its thermal
stability. Since the discovery of the standard thin accretion disk
(e.g., Shakura and Sunyaev 1973), many efforts have been made to
examine its various stability. When the mass accretion rate is
higher than a few percent of Eddington rate ($\dot{M}_{\rm Edd}
\equiv 10L_{\rm Edd}/c^2$), which roughly corresponds to $\ga 0.06
L_{\rm Edd}$, the innermost region of the disk will be
radiation-pressure-dominated. It was found that in this case the
disk will be both secularly (Lightman \& Eardley 1974) and thermally
(Shakura \& Sunyaev 1976; Piran 1978) unstable. Time-dependent
global numerical calculations (Honma et al. 1992; Szuszkiewicz \&
Miller 1998; Janiuk et al. 2002; Li et al. 2007) found that the
local thermal instability will result in the ``limit-cycle''
behavior. However, observations of the high/soft state of black hole
X-ray binaries have raised doubt to the above prediction (GD04). It
has been well established that the accretion flow in this state is
described by the standard thin disk model (Zdziarski \&
Gierli{\'n}ski 2004; Done et al. 2007). The luminosity of the
sources compiled in GD04 ranges from 0.01 to 0.5 $L_{\rm Edd}$, with
the highest one even exceeding the theoretical stable limit by
nearly one order of magnitude. Thus we should expect to see the
limit-cycle behavior. In contrast to this expectation, however,
observations show little variability which convincingly indicates
that they are thermally stable.

Some efforts have been made to solve this puzzle. For example, if a
large fraction of the dissipated energy is channeled into a corona
or outflow/jet the disk would be stable (Svensson \& Zdziarski
1994). However, the sources adopted in GD04 are all disk-dominated.
The absence of hard X-ray and radio emission directly rules out the
existence of corona or jet. Another idea is to assume that the
viscous stress is proportional to the gas pressure only, rather than
the sum of the gas and radiation pressure (Lightman \& Eardley 1974;
Stella \& Rosner 1984). As pointed out in GD04, the problem
with this idea is that the outcome of this modified viscosity is in
conflict with the observation that the color temperature correction
is constant for different accretion rates. Moreover, on the
theoretical side, in contrast to the above two ideas, numerical
simulations have shown that only a very small fraction of the
dissipated energy is channeled into the corona, which is consistent
with the absence of hard X-ray emission in high state of black hole
X-ray binaries, and the stress is proportional to the sum of gas and
radiation pressure (Hirose et al. 2006; Hirose et al. 2009a).
Recently, the issue of thermal instability has been addressed by the
radiation magnetohydrodynamic simulations of a vertically stratified
shearing box (Hirose et al. 2009a). Their results indicate that the
radiation-dominated disk is stable over $\sim 40$ cooling
timescales. They explained the stability as the lack of correlation
between the stress and the pressure within the cooling timescale.

In this paper, we revisit the thermal stability of thin disks
by linear analysis.
Different from previous analyses, we include the role of the
magnetic pressure and further assume that the magnetic field becomes
weaker when the gas temperature increases. Our analysis indicates
that the onset accretion rate for the instability becomes larger
compared with the previous result. The organization of our paper is as
follows. In \S2, we deliver the equations and conduct an analysis of
the thermal instability. The results are shown in \S3. The last
section (\S4) is devoted to a summary and discussion.

\section{Accretion disk model with toroidal magnetic fields}
\label{modeling}
\subsection{Basic equations}

The disk structure is described by the following equations.  The
equation of vertical hydrostatic equilibrium is
\begin{equation}
\frac{\partial{p}}{\partial z}+\rho \frac{\partial \psi}{\partial z}=0,
\end{equation}
where $p$ is the total pressure varying with $z$, $\rho$ is the mass
density, $\psi$ is the gravitational potential which is assumed to
be of the Newtonian form, i.e., $\psi(R,z)=-GM/\sqrt{R^2+z^2}$,
where $M$ is the mass of the black hole. Under the thin disk
approximation, we have $\partial p/\partial z\simeq -P_{\rm tot}/H$
and $\partial \psi/\partial z\simeq GMH/R^3$, where $H$ is the scale
height, and $P_{\rm tot}$ is the total pressure at the midplane.
Defining the surface density as $\Sigma \equiv 2\rho H$, we have
\begin{equation}
P_{\rm tot}=P_{\rm gas}+P_{\rm rad}+P_{\rm mag}=\frac{GM\Sigma H}{2R^3}.
\end{equation}
Here the gas pressure $P_{\rm gas}=2k_{\rm B}\rho T/m_{\rm H}$,
where $k_{\rm B}$ is the Boltzmann constant, $m_{\rm H}$ is the
hydrogen mass, and $T$ is the temperature at the midplane. The
radiation pressure $P_{\rm rad}=aT^4/3$, where $a$ is the radiation
constant. Since in the innermost region of the accretion flow the
magnetic field is dominated by its azimuthal component
$B_{\varphi}$, the magnetic pressure $P_{\rm
mag}=B_{\varphi}^2/8\pi$.

The energy equation is written as the balance between the viscous
heating and the cooling through radiation and advection, i.e.,
\begin{equation}
Q_{\rm vis}^+=Q_{\rm rad}^-+Q_{\rm adv}^-.
\end{equation}
The vertically integrated viscous heating rate is
\begin{equation}
Q_{\rm vis}^+=-T_{r\varphi}R\frac{d\Omega}{dR},
\end{equation}
where $T_{r\varphi}$ is the stress (see eq. 8 below), and
$\Omega=\Omega_{\rm K}$ with $\Omega_{\rm K}\equiv\sqrt{GM/R^3}$ is
the Keplerian angular velocity. The radiative cooling rate is
\begin{equation}
Q_{\rm rad}^-=\frac{32\sigma T^4}{3\tau},
\end{equation}
where $\tau=\kappa \Sigma/2$ is the vertical optical depth,
$\kappa\simeq 0.4~{\rm cm^2~g^{-1}}$, since the opacity is dominated
by the electron scattering in the inner region of the disk, and
$\sigma$ is the Stefan-Boltzmann constant. The advective cooling
term is given by (Abramowicz et al. 1995)
\begin{equation}
Q_{\rm adv}^- =\frac{\xi \dot{M}\Omega_{\rm K}^2 H^2}{2\pi R^2}.
\end{equation}
We set $\xi=1.5$ in the present work. Our calculations indicate that
the role of advection term is negligible in term of the stability
analysis of the radiation-dominated thin disk. This term is included
for the discussion of the thermal equilibrium solutions in \S3.2,
which contain optically thick advection-dominated accretion disks
(or slim disks; Abramowicz et al. 1988).

The angular momentum conservation equation is expressed as
\begin{equation}
\dot{M}(\Omega_{\rm K}R^2-l_{\rm in})=2\pi R^2 T_{r\varphi},
\end{equation}
where $l_{\rm in}=\sqrt{GMR_{\rm in}}$ is the specific angular momentum at
the inner edge of the disk, and $R_{\rm in}$ is fixed to be $3R_{\rm g}$,
with $R_{\rm g}\equiv 2GM/c^2$ being the Schwarzschild radius.

It is now widely believed that the MHD turbulence associated with
the magneto-rotational instability (MRI) is the main mechanism for
the angular momentum transfer in accretion disks (e.g., Balbus \&
Hawley 1998). MHD simulations have shown that, the stress accounting
for the angular momentum transport is dominated by the Maxwell
stress, rather than the Reynolds stress (Hawley et al. 1995).
Besides, we have also learnt from the simulations that the ratio of
the Maxwell stress to the total pressure nearly maintains constant
(Hawley \& Krolik 2001; Machida et al. 2006; Pessah et al. 2007;
Hirose et al. 2009a). We therefore have
\begin{equation}
T_{r\varphi}=2\alpha P_{\rm tot}H ,
\end{equation}
where $\alpha$ is the viscous parameter.

To close the set of equations, we need one more constraint, which is
to connect the magnetic field with other physical quantities. As the
thermal instability is mainly concerned in this study, here, we
relate the response of the magnetic field to the perturbation of the
scale height in a linearized fashion, i.e.,
\begin{equation}
\frac{\delta B_\varphi}{B_\varphi}=-\gamma \frac{\delta H}{H} ,
\end{equation}
where $\gamma$ is a free parameter. The integrated form of eq. (9)
is $\Phi_\gamma \equiv B_\varphi H^\gamma={\rm constant}$. We assume
that $\gamma>0$, i.e., the magnetic field will become weaker with an
increase of height (or temperature). This assumption is supported by
the MHD numerical simulation of Machida et al. (2006) where they
found that the magnetic field becomes stronger when the disk shrinks
vertically, although we note that their simulation is for a hot
accretion flow rather than a thin disk. We will discuss the physical
meaning of $\Phi_\gamma$ in \S4. With the above equations we can
obtain a thermal equilibrium solution at a certain radius for given
$M$, $\dot{M}$, $\alpha$ and $\Phi_\gamma$ (\S3.2).

\subsection{Thermal instability analysis}

Since the thermal instability timescale is much shorter than the viscous
timescale, the surface density $\Sigma$ is taken to be constant. Defining
$\beta_{\rm gas}\equiv P_{\rm gas}/P_{\rm tot}$ and $\beta_{\rm mag}\equiv
P_{\rm mag}/P_{\rm tot}$, we can obtain from eqs. (2), (4)-(6) that
\begin{eqnarray}
d\ln P_{\rm tot}
&&=d\ln H \nonumber \\
&&=\beta_{\rm gas}(d\ln T-d\ln H)+4(1-\beta_{\rm gas}-\beta_{\rm
mag}) d\ln T+2\beta_{\rm mag}d\ln B_\varphi ,
\end{eqnarray}
and
\begin{eqnarray}
d\ln Q_{\rm vis}^+-d\ln(Q_{\rm rad}^-+Q_{\rm adv}^-)=d\ln T_{r\varphi}-4(1-f_{\rm adv}) d\ln T+f_{\rm
adv}(d\ln \dot{M}+2 d\ln H) ,
\end{eqnarray}
where the advection factor $f_{\rm adv}$
is defined as $f_{\rm adv}\equiv Q_{\rm adv}^-/(Q_{\rm rad}^-+Q_{\rm adv}^-)$.
Eqs. (7)-(9) give that
\begin{equation}
d\ln \dot{M}=d\ln T_{r\varphi}=d\ln P_{\rm tot}+d\ln H ,
\end{equation}
and
\begin{equation}
d\ln B_\varphi=-\gamma d\ln H .
\end{equation}
Combining eqs. (10) and (13), we have
\begin{equation}
d\ln P_{\rm tot}=d\ln H=\frac{4-3\beta_{\rm gas}-4\beta_{\rm mag}}{1+\beta_{\rm gas}+2\gamma\beta_{\rm mag}}d\ln T .
\end{equation}

Substituting eqs. (3), (12) and (14) into eq. (11), we finally get
\begin{eqnarray}
\left[\frac{\partial(Q_{\rm vis}^+-Q_{\rm rad}^--Q_{\rm
adv}^-)}{\partial T}\right]_{\Sigma}
 \frac{T}{Q_{\rm vis}^+}
= \frac{2-5\beta_{\rm gas}-4(1+\gamma)\beta_{\rm mag}-6f_{\rm adv}+
8f_{\rm adv}\beta_{\rm gas}+(8+4\gamma)f_{\rm adv}\beta_{\rm mag}}
{1+\beta_{\rm gas}+2\gamma\beta_{\rm mag}} .
\end{eqnarray}
Since the thermal instability condition is $[\partial(Q_{\rm
vis}^+-Q_{\rm rad} ^--Q_{\rm adv}^-)/\partial T]_{\Sigma}>0$, and
the denominator of eq. (15) is always positive when $\gamma>0$, then
the thermal instability criterion can be taken as
\begin{eqnarray}
\Delta=2-5\beta_{\rm gas}-4(1+\gamma)\beta_{\rm mag}-6f_{\rm adv}+8f_{\rm adv}
\beta_{\rm gas}+(8+4\gamma)f_{\rm adv}\beta_{\rm mag}>0 .
\end{eqnarray}
For the cases without magnetic field, i.e., $\beta_{\rm mag}=0$, the
above criterion is reduced to (Gu \& Lu 2007):
$$2-5\beta_{\rm gas}-6f_{\rm adv}+8\beta_{\rm gas} f_{\rm adv} > 0.$$

\section{Results}
\label{results}

\subsection{Stability of disks with different magnetic field strength}

\begin{figure*}
\plottwo {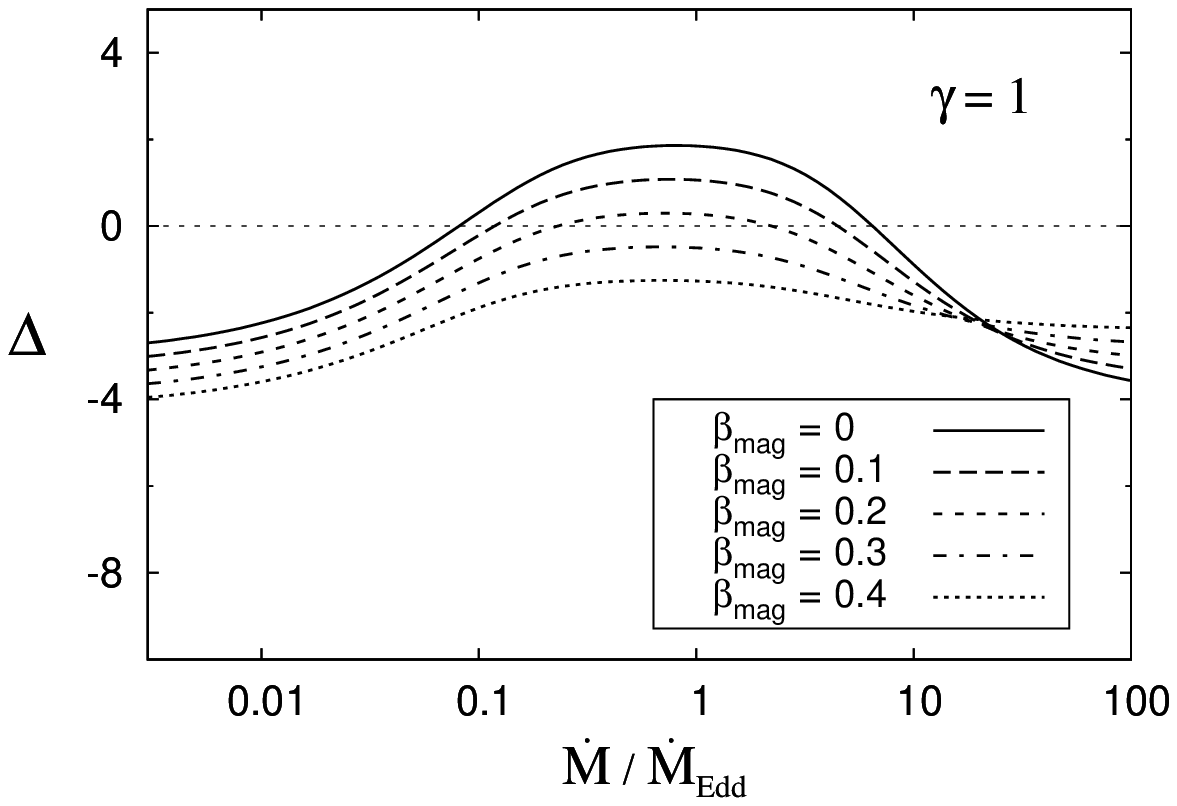}{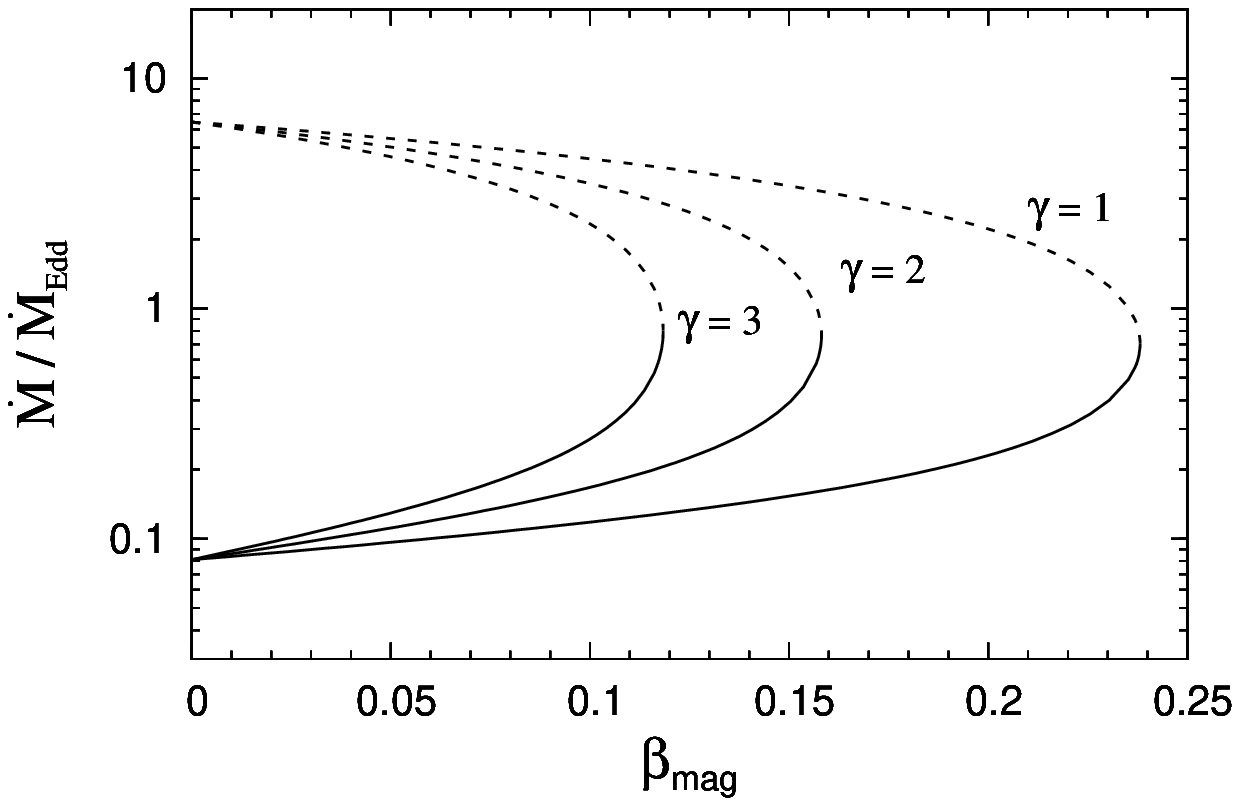}
\caption{The thermal stability of the disk at $R=10R_{\rm g}$ for various
$\dot{M}$ and $\beta_{\rm mag}$. $M=10M_\odot$ and $\alpha=0.1$ are adopted.
The left panel shows the variation of $\Delta$ with $\dot M$ for given
values of $\beta_{\rm mag}$, where $\gamma$ is set to be 1.
The solid line corresponds to the none magnetic field cases.
The right panel shows the variation of the lower and upper critical
accretion rates ($\Delta=0$) with $\beta_{\rm mag}$. Each of the
curves corresponds to a specific value for $\gamma$.}
\end{figure*}

We calculate the thermal equilibrium solution of the above
equations. The left panel of Fig. 1 shows the value of $\Delta$ at
$R = 10 R_{\rm g}$ for various $\dot{M}$ and $\beta_{\rm mag}$. The
value of $\gamma$ is selected to be 1 as an illustration. We can see
that when $\beta_{\rm mag} \la 0.24$, i.e., the magnetic field is
relatively weak, there exist two critical mass accretion rates
corresponding to $\Delta = 0$ for each given $\beta_{\rm mag}$. When
$\beta_{\rm mag} \ga 0.24$, we always have $\Delta<0$. That means
the disk will be thermally stable for any $\dot M$. The results are
similar for the cases of other $\gamma$. The right panel of Fig. 1
shows the variation of the two critical accretion rates with
$\beta_{\rm mag}$, where the solid and dashed lines correspond to
the lower and upper critical rates, respectively. The lower critical
rate is of interest to us since it is relevant to the thin disks. It
corresponds to the threshold of the accretion rate above which the
disk is unstable. The upper one corresponds to slim disks with
$f_{\rm adv} \sim 1/3$ (see \S3.2). For each $\gamma$, the region
inside the corresponding parabolic curve denotes the thermally
unstable solutions, while the region outside corresponds to the
stable ones. The solid lines in the right panel illustrate that the
threshold of accretion rate above which the disk is thermally
unstable increases with the ratio of magnetic pressure to total
pressure. Also, we find that if $\gamma$ takes a larger value, i.e.,
the magnetic field reacts more intensely according to a perturbation
in the scale height, the critical rate increases more significantly
with $\beta_{\rm mag}$, and thus a smaller cutoff of the possible
$\beta_{\rm mag}$ for thermal instability appears.

We interpret the above result as follows. Suppose there is a small
increase in the temperature $T$, the radiative cooling will increase
following eq.~(5), while the response of viscous heating to
temperature is not so simple. As the temperature increases, the
radiation pressure and further the scale height increase
accordingly, which will result in the increase of viscous heating
(ref. eqs. 4 and 8). The key point is that the magnetic field
$B_\varphi$ will decrease (eq. 9), and this effect appears more
important when $\gamma$ is relatively large. Therefore, compared
with the case without magnetic field, the increase of the total
pressure and thus the scale height will become less significant. In
other words, the inclusion of magnetic pressure will result in a
weaker dependence of viscous heating on temperature, thus the
magnetic field can conspicuously suppress the thermal instability.

\subsection{Local thermal equilibria}

\begin{figure}
\plotone{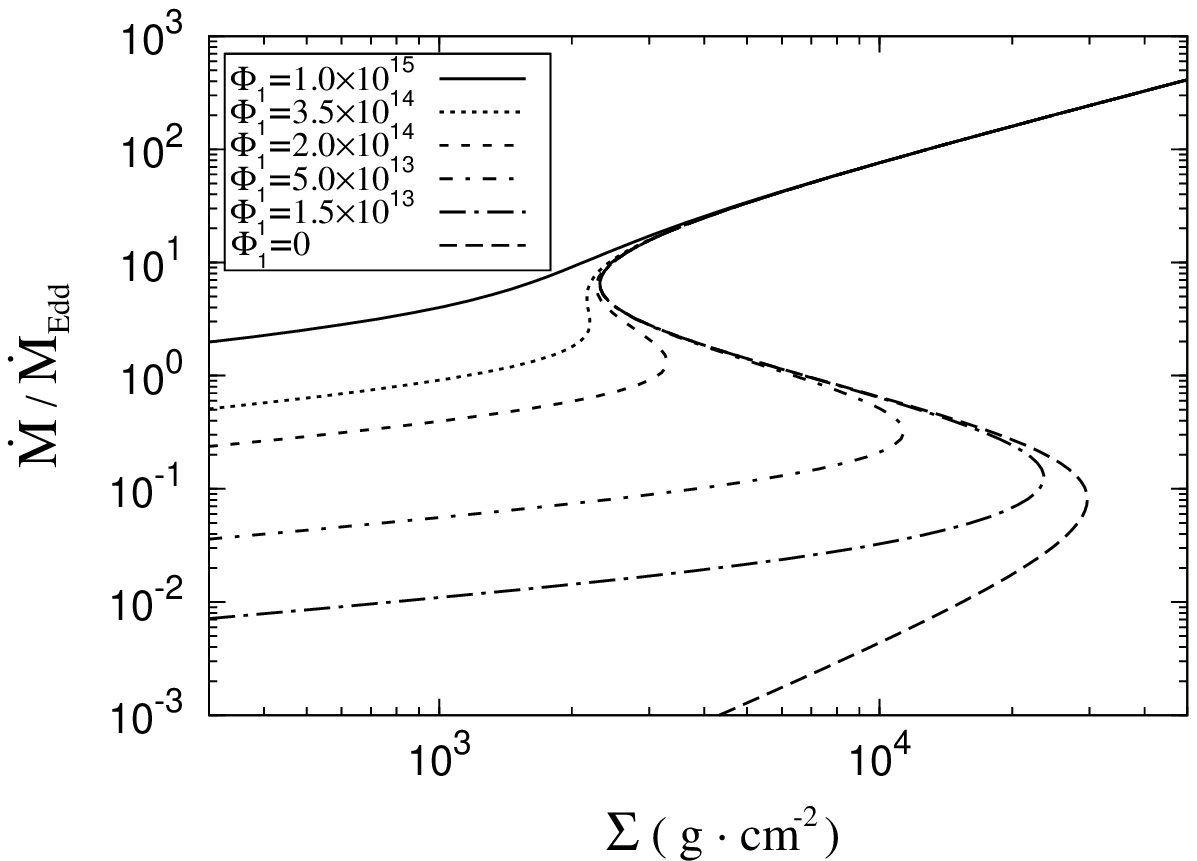}
\caption{The thermal equilibrium
curves of the thin disk at $10 R_{\rm g}$ for different $\Phi_1$, which is defined
as $\Phi_1 \equiv B_\varphi H$ and is in unit of Gs$\cdot$cm.
Other model parameters are $M=10M_{\odot}$ and $\alpha=0.1$~.}
\end{figure}

It is convenient to plot an $\dot{M} \sim \Sigma$ diagram of local
disk solutions to directly show the properties of disk stability.
For simplicity, we also select the case of $\gamma=1$ as
an instance, which is presented in Figure 2.
On those curves in this figure, a negative slop indicates that the
solution is thermally unstable. We can see that the threshold of accretion
rate for thermal instability, corresponding to the lower turning
points of each curve, rises accordingly with increasing $\Phi_1$, which is in
agreement with Fig.~1. For the dotted and the short
dashed curves, the thresholds of accretion rate
are all $\sim \dot{M}_{\rm Edd}$, roughly one order of magnitude
larger than the case without magnetic field (the long dashed curve),
which is supposed to explain the observed
high luminosity of {\em thermally stable} high/soft state of some
black hole X-ray binaries (GD04). While for the solid curve, as the
corresponding $\Phi_1$ is large enough, the two turning points
both disappear.

All the curves in Fig. 2 share the same upper branch. That is
because in the cases of large accretion rates, i.e., $\ga10
\dot{M}_{\rm Edd}$, the advective cooling becomes important and the
radiation pressure turns to be very large. Compared with the large
radiation pressure, the magnetic pressure can be neglected (the
value of $P_{\rm mag}/P_{\rm rad}$ is of the order of $\sim 10^{-2}$
at the upper turning point of each curve except the solid one
whereas $f_{\rm adv}$ approaches 1/3, and decreases rapidly as the
accretion rate increases). Hence, the effect of magnetic field is
negligible, and all the curves with different $\Phi_1$ approach the
slim disk limit. It is qualitatively the same when $\gamma$ takes
other values. We would like to emphasize that there is no
requirement that the disk solution would trace the solution curves
in Figure 2 when the increase of the accretion rate is adequately
slow. In fact, the conservation of $\Phi_\gamma$ may not hold beyond
a thermal timescale.

\section{Summary and discussion}

Previous analysis shows that the radiation-dominated standard thin
disk is thermally unstable when $L\ga 0.06 L_{\rm Edd}$. This
conclusion, however, is in conflict with observations of, e.g., the
high state of black hole X-ray binaries. Most previous analyses
neglect the role of magnetic pressure. In this paper, by taking the
magnetic pressure into account, we have revisited the thermal
stability of the radiation-dominated standard thin disk. We have
derived a general criterion for the thermal instability (eq. 16). We
find that the threshold of accretion rate above which the solution
becomes unstable increases significantly with the increasing ratio
of magnetic pressure and total pressure, $\beta_{\rm mag}$ (refer to
Fig. 1). If $\beta_{\rm mag}$ is large enough, say $\beta_{\rm
mag}\ga 0.24$ for $\gamma=1$ or $\beta_{\rm mag}\ga 0.12$ for
$\gamma=3$, our model predicts that the disk will be stable for any
$\dot{M}$ (again refer to Fig. 1). We do not favor this result since
the required $\beta_{\rm mag}$ is significantly larger than the
typical value of $\beta_{\rm mag}\sim 0.1$ given by general
numerical simulations. For $\beta_{\rm mag}\sim 0.1$, from Fig. 1 we
expect that the thermal stability of the high state of black hole
X-ray binaries with luminosity as high as $0.5 L_{\rm Edd}$ can be
explained if $\gamma \ga 3$\footnote{Throughout the paper we set
$\alpha=0.1$. If the value of $\alpha$ is smaller, the required
$\gamma$ can be smaller to explain the observed stable high state.}.

The key assumption in our analysis is that during the thermal
perturbation the changes of magnetic field and of scale height
satisfy a constraint described by eq. (9), namely, {\em the magnetic
field $B_{\varphi}$ will become weaker with the increase of
temperature (or equivalently scale height $H$)}. The parameter
$\gamma$ in eq. (9) denotes how strong the response of the magnetic
field is to a thermal perturbation. Therefore, compared with the case
of not including the magnetic pressure, the increase of total
pressure and further the scale height and viscous heating with the
temperature become weaker (ref. eqs. 2, 4 \& 8). This is the reason
why the disk tends to be thermally stable.

Unfortunately we are unclear about the value (or the range) of
$\gamma$ because of the complexity of processes such as MRI, dynamo,
and reconnection which can strengthen or weaken the magnetic field
within timescales shorter than the thermal one. One extreme case is
that all these processes are turned off thus the magnetic field is
frozen in the accretion flow. In this case, we obviously have
$\gamma=1$, and $\Phi_1=B_{\varphi}H$ represents half of the
toroidal magnetic flux per unit radius. The conservation of $\Phi_1$
is exactly what has been adopted in the thermal stability analysis
of a magnetically dominated ``low-$\beta$'' disk (Machida et al.
2006). In another work studying this type of disk (Oda et al. 2010),
the advection rate of toroidal magnetic flux ($\equiv 2v_rB_\varphi
H$, where $v_r=\dot{M} /2\pi r\Sigma$ is the radial velocity of the
accretion flow) is assumed to be constant during a thermal
perturbation. This corresponds to $\gamma=3$, which is because from
eqs. (2), (7) and (8), we can deduce that $v_r \propto H^2$ and
therefore $B_\varphi\propto H^{-3}$. We also notice that the model
constructed in Oda et al. (2009) corresponds to $\gamma=2$. In this
case, the conservation of $\Phi_2=B_{\varphi}H^2$ comes from the
specific configuration of both the magnetic field and the outer
boundary condition, which does not hold a simple physical meaning.
In spite of the uncertainty in the value of $\gamma$, we believe
that the positive sign of $\gamma$, or equivalently the negative
response of $B_{\varphi}$ to $T$, should capture the real physics
and have essential influence on the thermal stability of
radiation-dominated thin disks. It will be interesting to
investigate the value of $\gamma$ by detailed numerical simulations.
On the other hand, it is noteworthy that the reversal of toroidal
magnetic field was found by vertically stratified shearing box
simulations of thin disks (e.g., Brandenburg et al. 1995; Stone et
al. 1996), and most recently by global simulations (O'Neill et
al. 2010). This phenomenon was reported to occur on $\sim$ 10 orbit
timescales, comparable to the thermal timescale. If this phenomena
is coupled with the thermal perturbation process, which is unclear
to us, it will be a problem for our model, since it will be in
conflict with our assumption of the conservation of $\Phi_\gamma$\footnote
{In this context we note that our model actually only
requires the negative response of the {\em absolute value of
$B_{\varphi}$} to temperature. But even so, if the field reversal is
coupled with thermal perturbation process, since the timescales of
the two processes are comparable, our assumption of eq. (9) will
still be incorrect.}. Another concern comes when we try to
explain the stability of the radiation-dominated simulations
presented in Hirose et al. (2009b) based on our model. We find that
it requires a rather large value of $\gamma$ to explain some of
their simulations. For example, their model 0519b, the most
radiation-dominated one, requires $\gamma\ga 7$. Such a high value
feels uncomfortable. Of course, it may be too demanding to require a
simple one-dimensional analytical model to completely explain
three-dimensional MHD simulations.

Our model predicts that if the magnetic pressure
is not too strong, e.g., $\beta_{\rm mag} \la 0.24$ for $\gamma=1$,
the sources with accretion rates higher than the
threshold should still manifest thermal instability.
We note in this context that GRS 1915+105 may be one evidence for
our prediction. Different from other sources, GRS 1915+105 holds
strong variability which seems to be well interpreted as the thermal
instability of a radiation-dominated thin disk (e.g., Belloni et al.
1997; Janiuk et al. 2002). GD04 speculated that
the reason for the uniqueness of this source might be that it goes
to higher luminosity (or equivalently, accretion rate) than all
other sources in their sample. Actually GRS1915+105 is likely to be
super-Eddington (Done et al. 2004). This prediction could
discriminate our model from others such as the one presented in
Hirose et al. (2009a).

\begin{acknowledgements}

We thank the referee, Omer Blaes, for useful communications and his
beneficial comments. This work was supported by the National Basic
Research Program of China under grant 2009CB824800, the National
Natural Science Foundation of China (grants 10821302, 10825314,
10833002, and 11073015), and the CAS/SAFEA International Partnership
Program for Creative Research Teams.

\end{acknowledgements}

\end{document}